# Skeletal Density Measurements for Adsorbent Nanomaterials

Andrew K. Gillespie[a]*, Zeke A. Piskulich[b], Ernest Knight[c], Matthew Prosniewski[d], and Peter Pfeifer[e]

**AFFILIATIONS**

[a]Department of Physics and Astronomy, Texas Tech University, Lubbock TX 79409, USA

[b]Department of Chemistry, Boston University, Boston, MA 02215, USA

[c]Euclid Techlabs LLC, Bolingbrook, IL 60440, USA

[d]Ally Financial, Detroit, MI 48226, USA

[e]Department of Physics and Astronomy, University of Missouri, Columbia MO 65201, USA

*Author to whom correspondence should be addressed: a.gillespie@ttu.edu

**ABSTRACT**

Due to their tunable material properties, sorptive materials have a wide range of applications in energy storage, water treatment, carbon capture, analytical chemistry, and more. One crucial factor in determining the effectiveness of such materials is their skeletal density, or "true density" because it is often used to calculate key metrics, such as storage capacities. In this paper, we present skeletal density measurements through helium pycnometry for several types of adsorbent carbon materials derived from either corncob, sawdust, coffee grounds, polyvinylidene chloride (PVDC), graphitic carbon nitride (GCN), or metal organic frameworks (MOFs). The measured skeletal density of sawdust-based activated carbon was $2.02 \pm 0.05$ g/cm$^3$. The measured skeletal density of coffee-based activated carbon was $2.23 \pm 0.06$ g/cm$^3$. We also expound upon the impact that skeletal density has upon hydrogen excess adsorption measurements and other calculated engineering quantities. If a skeletal density is underestimated by 10%, it can affect the room temperature excess adsorption by at least 5% at 100 bar and by 7% at 200 bar, depending on the material type.

## 1. INTRODUCTION

The United States Department of Energy recently announced the selection of seven hydrogen hub projects across the country aimed at improving clean hydrogen production, delivery, and storage.[1-4] Hydrogen volumetric storage near room temperature is achieved through compression and can be improved using an appropriate adsorbent material. Adsorption is the process by which an adsorbate gas increases in density when in the vicinity of a material interface due to Van der Waals interactions. Among several variables, this effect is largely due to the extent of the interfacial area, which is generated by increasing the network of channels in the pore volume of the adsorbent material. Though adsorption instruments vary greatly in type and technique, they all measure the same quantity: excess adsorption.

For high surface area adsorbent materials, excess adsorption is one of the only directly measurable quantities without theoretical assumptions.[5] Excess adsorption may be thought of as the mass of the adsorbed film minus the mass of an equal volume of compressed gas.

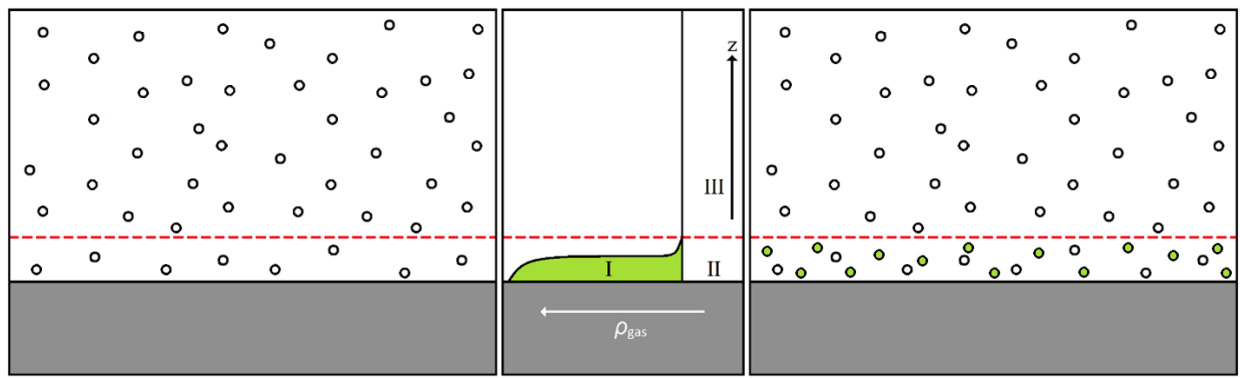

**Figure 1:** An adsorption schematic of gas densities versus distance from the adsorbent surface. *Left*: Volume of compressed gas at a constant density in the absence of an adsorbing potential. *Right*: Compressed gas in the presence of an adsorbing potential. The gas densifies near the adsorbing surface. Colored molecules have been adsorbed in excess of those that would be present in the absence of an adsorbing potential. Thus, they are defined as the excess adsorption. *Middle*: Graphical representation of gas density vs. distance from the adsorbent surface. The three regions indicated show excess adsorption (I), absolute adsorption (I + II), and total storage capacity (I + II + III).[6,7]

Though excess adsorption is the only quantity that may be directly measured, storage capacities and absolute adsorption are derived from the excess adsorption measurement using assumptions about the pore volume, adsorbed film thickness, or surface geometry.

Volumetric methods for determining excess adsorption have been used for almost 90 years. Most modern volumetric sorption instruments employ the Sieverts' method, monitoring changes in pressure and temperature in order to measure excess adsorption. Basic manometric instruments consist of two known volumes, referred to here as the dosing volume, $V_d$, and the reactor volume, $V_r$, separated by a valve. $V_{sk}$ is the skeletal volume of the sample [8].

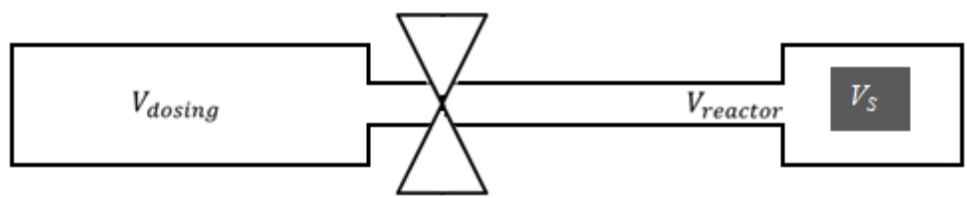

**Figure 2:** Manometric instrument consisting of two known volumes, separated by a valve.

The dosing and reactor volumes can easily be calibrated with an uncertainty of less than 0.5% using a NIST-traceable volume standard or using silicon displacer beads.[9,10] If $V_d$, $V_r$, and

$V_{sk}$ are known, manometric instruments may use the conservation mass to perform sorption measurements. Beginning with an evacuated system, the dosing volume is pressurized with the adsorbate gas. The temperature and pressure are measured and the mass density, $\rho_1$, is calculated using an appropriate equation of state.[11-13] The mass of adsorbate gas contained in the system is now given by **Equation 1**.

$$m_g = \rho_1 V_d \quad (1)$$

$m_g$ is the mass of the adsorbate gas, $\rho_1$ is the density of the gas, and $V_d$ is the dosing volume. After an adequate equilibration time, the valve separating the dosing volume and reactor volume is opened allowing the adsorbate gas to enter the reactor volume. Again, equilibrium is reached and the gas density, $\rho_2$, is calculated based on the new pressure and temperature measurements. In the presence of a non-adsorbing sample, the mass of the adsorbate gas may now be expressed as

$$m_g = \rho_2(V_d + V_r - V_{sk}) \quad (2)$$

where $V_{sk}$ is the skeletal volume of the adsorbent material. This volume may be expressed in terms of mass and density as $V_{sk}=m_s/\rho_{sk}$ in applications where it is more convenient to use the skeletal density. In the presence of an adsorbing sample, the mass of adsorbate gas in the system may be expressed in terms including the adsorbed film volume, $V_{film}$, and the density of the gas in the adsorbed phase, $\rho_{film}$, as shown in **Equation 3**.

$$m_g = \rho_2(V_d + V_r - V_{sk}) + V_{film}(\rho_{film} - \rho_2) \quad (3)$$

The last term in this expression is the definition of excess adsorption. Thus, the mass of adsorbate gas contained in the system may be expressed as

$$m_g = \rho_2(V_d + V_r - V_{sk}) + m_{exc} \quad (4)$$

If two or more subsequent adsorption measurements are performed, then the expression expands to account for the gas amounts remaining in each volume.[14,15] For this paper, we will focus on singular excess adsorption measurements as they pertain to the effect of skeletal density. Setting **Equation 1** equal to **Equation 4** yields a simple expression for excess adsorption.

$$m_{exc} = V_d(\rho_1 - \rho_2) - \rho_2(V_r - V_{sk}) = V_d(\rho_1 - \rho_2) - \rho_2\left(V_r - \frac{m_s}{\rho_{sk}}\right) \quad (5)$$

Other derived engineering quantities, such as gravimetric storage capacity and volumetric storage capacity are heavily dependent upon the skeletal volume [16,17]. The skeletal volume may be determined by taking headspace measurements using helium pycnometry. These measurements are performed in a manner similar to any excess adsorption measurement, but using a non-adsorbing gas. These measurements are highly sensitive to uncertainties in dosing and reactor volumes. Therefore, helium pycnometry measurements should be performed in a well-calibrated, voluminous reactor and on a large amount of the adsorbent sample to reduce the relative error.

Once the system is fully calibrated such that the dosing volume, $V_d$, and the reactor volume, $V_r$, are known, one can begin measuring the sample volume. Setting **Equation 1** equal to **Equation 2** and solving for the skeletal volume yields the following:

$$V_{sk} = V_d + V_r - \frac{\rho_1 V_d}{\rho_2} \tag{6}$$

To obtain the skeletal density, simply take a mass measurement and divide it by the skeletal volume. In this paper, we present measurements of the skeletal densities of several types of adsorbent materials, as well as an uncertainty analysis on their effect on the derived quantities of storage capacities.

## 2. Instrumentation and Methods

A volumetric method was used to determine skeletal volumes and excess adsorption by monitoring changes in the pressure and temperature of the system. Helium pycnometry and excess adsorption isotherms were measured using a modified[15] HTP-1 Volumetric Analyser manufactured by Hiden Isochema.[18] The HTP-1 is a manometric instrument capable of measurements at pressures ranging up to 200 bar using a Baratron manometer. The HTP-1 has a low-pressure mode that was used for these helium pycnometry measurements. Through hundreds of calibration measurements, the dosing volume was determined to be $51.75 \pm 0.02$ cm$^3$ and the reactor volume was determined to be $6.32 \pm 0.02$ cm$^3$. Helium, an inert gas, was used to determine the headspace volume available to the gas followed by excess adsorption measurements.

## 3. Results

As an example, the skeletal volume and skeletal density of one sample is shown in **Figure 3**. This sample was named 2.3K-0810 and was derived from waste coffee grounds.

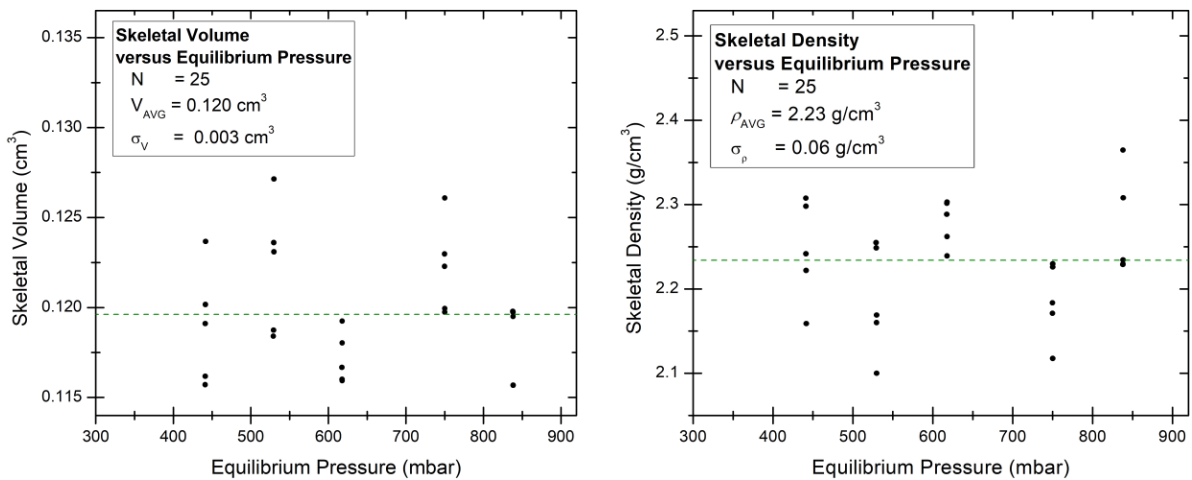

**Figure 3.** Helium pycnometric results for the coffee-based activated carbon.

**Figure 3** shows this sample has a skeletal density of $2.23 \pm 0.06$ g/cm$^3$. The reported uncertainty is based on the standard deviation in 25 measurements of the skeletal density. After the headspace volume was determined, measurements of excess adsorption, gravimetric storage capacity, and volumetric storage capacity may be performed. The skeletal volumes and densities for several types of adsorbent materials are shown in **Table 1**. Uncertainties in skeletal density are reported as the standard deviation across repeat measurements.

**Table 1:** Skeletal volumes and densities of various adsorbent materials.

| Sample Name | Precursor or Material Type | N | Pressure Range [mbar] | Skeletal Volume [cm$^3$] | Mass [g] | Skeletal Density [g/cm$^3$] |
|---|---|---|---|---|---|---|
| 1.85K-0099 | Corncob | 41 | 260-800 | 0.139 | 0.350 | 2.51±0.06 |
| 2.3K-0810 | Coffee Grounds | 25 | 440-840 | 0.120 | 0.267 | 2.23±0.06 |
| 2.5K-0753 | Corncob | 10 | 530-800 | 0.107 | 0.244 | 2.28±0.04 |
| 3K-0285 | Corncob | 34 | 260-840 | 0.110 | 0.251 | 2.28±0.09 |
| 4K-0749 | Corncob | 20 | 530-800 | 0.085 | 0.178 | 2.09±0.06 |
| 4K-0803 | Corncob | 20 | 260-800 | 0.071 | 0.166 | 2.33±0.09 |
| 5K-0812 | Corncob | 10 | 530-800 | 0.064 | 0.140 | 2.21±0.10 |
| MWV-0260 | Sawdust | 17 | 800 | 0.109 | 0.277 | 2.54±0.03 |
| HKUST-1 (MOF-199) [19] | Cu MOF | 30 | 90-800 | 0.138 | 0.257 | 1.87±0.09 |
| GCN-1001 [20] | Melamine | 19 | 90-800 | 0.289 | 0.549 | 1.90±0.02 |
| GCN-1002 [20] | Melamine | 23 | 340 | 0.666 | 1.155 | 1.74±0.14 |
| GCN-1003 [20] | Melamine | 30 | 800 | 0.079 | 0.139 | 1.76±0.04 |
| GCN-1008 [20] | Melamine | 15 | 2000-18000 | 0.126 | 0.247 | 1.96±0.09 |
| GCN-1010 [20] | Melamine | 20 | 90-800 | 0.250 | 0.463 | 1.85±0.04 |
| GCN-1013-e [20] | Melamine | 14 | 260-800 | 0.020 | 0.026 | 1.31±0.17 |
| GCN-1026-e [20] | Melamine | 30 | 260-800 | 0.097 | 0.223 | 2.30±0.08 |
| GCN-1039 [20] | Melamine | 60 | 260-840 | 0.322 | 0.601 | 1.86±0.02 |
| GCN-1040 [20] | Melamine | 38 | 440-840 | 0.224 | 0.418 | 1.87±0.01 |
| GCN-1067-e [20] | Melamine | 40 | 350-800 | 0.286 | 0.467 | 1.63±0.02 |
| GCN-1200-Si | Melamine + Ludox HS-40 | 30 | 350-840 | 0.326 | 0.743 | 2.27±0.05 |
| XF-GCN-1141 | Melamine | 25 | 440-840 | 0.074 | 0.114 | 1.56±0.14 |
| xGnP-C-500 [21,22] | Exfoliated Graphite | 15 | 620-840 | 0.178 | 0.397 | 2.23±0.08 |
| xGnP-C-750 [21,22] | Exfoliated Graphite | 30 | 350-840 | 0.143 | 0.280 | 1.96±0.06 |
| Cabot-EXP-14008 [23,24] | | 15 | 260-800 | 0.176 | 0.363 | 2.06±0.05 |
| Cabot-EXP-14009 [23,24] | | 10 | 530-800 | 0.086 | 0.181 | 2.11±0.05 |
| BR-0311 | Sawdust | 39 | 260-840 | 0.130 | 0.263 | 2.02±0.05 |
| PVDC-0414 | PVDC | 5 | 800 | 0.246 | 0.527 | 2.15±0.03 |
| PVDC-0423 | PVDC | 24 | 260-800 | 0.311 | 0.541 | 1.74±0.01 |
| Soot-1b | Sawdust | 29 | 440-850 | 0.048 | 0.109 | 2.27±0.17 |

Skeletal densities of activated carbon materials have been reported to range from 1.8 – 2.5 g/cm$^3$. The powdered activated carbon materials within this study are consistent with that range. Those with higher densities likely contain inclusions of metals from the synthesis reaction vessel.

MSC-30, a highly studied carbon adsorbent, also falls within this range with a reported skeletal density between 2.0 – 2.2 g/cm$^3$, close to that of graphene.[16,25-33] The measured skeletal density of HKUST-1 was consistent with the published value in work by Britt et al.[19] The skeletal density of non-exfoliated PVDC-based adsorbent materials was slightly lower at 1.74 g/cm$^3$. Since PVDC was used as a binder for the briquetted material BR-0311, the skeletal density of the finished material was reduced to about 2.02 g/cm$^3$ compared to the higher skeletal density of the precursor powder, MWV-0260. Non-exfoliated GCN materials had skeletal densities consistently around 1.85 g/cm$^3$ whereas the skeletal densities of GCN materials that underwent additional treatment procedures varied from 1.6 – 2.3 g/cm$^3$ depending on the synthesis method, as outlined by Stalla.[20]

## 4. Uncertainty Analysis for Excess Adsorption and Storage Capacities

Other considerations should be given to the skeletal density of adsorbent materials. The exhaustive nature of accurate helium pycnometry measurements, the typically small sample size of experimental adsorbent material, and the high throughput of adsorption measurements often make it impractical for laboratories to invest sufficient time to determine skeletal densities for all samples with a high degree of accuracy. For that reason, many research groups (1) take a single headspace measurement or (2) find it reasonable to determine the skeletal density of representative materials and apply it to all materials within that subclass being screened. However, even small uncertainties in the dosing and reactor volumes will influence the excess adsorption. Consequently, any uncertainties or assumptions used will compound when calculating engineering quantities, such as gravimetric and volumetric storage capacities.

Assuming separable variables, the uncertainty in excess adsorption measurements may be calculated based on $V_d$, $V_r$, and $V_{sk}$ using the variance formula, as shown in **Equation 7**.[34,35]

$$\sigma_{mexc} = \sqrt{\left(\frac{dm_{exc}}{dV_{sk}}^2 \sigma_{Vsk}^2\right) + \left(\frac{dm_{exc}}{dV_d}^2 \sigma_{Vd}^2\right) + \left(\frac{dm_{exc}}{dV_r}^2 \sigma_{Vr}^2\right)} \qquad (7)$$

$$= \sqrt{(\rho_2^2 \sigma_{Vsk}^2) + ((\rho_1 - \rho_2)^2 \sigma_{Vd}^2) + (\rho_2^2 \sigma_{Vr}^2)}$$

Skeletal densities are often used in determining the gravimetric excess adsorption measurements for high throughput laboratories. In cases where it is more convenient to calculate these quantities based on the sample mass and skeletal density, the uncertainty in excess adsorption is represented by **Equation 8**.

$$\sigma_{mexc} = \sqrt{\left(\frac{\rho_2 m_s^2}{\rho_{sk}^2} \sigma_{\rho sk}^2\right) + ((\rho_1 - \rho_2)^2 \sigma_{Vd}^2) + (\rho_2^2 \sigma_{Vr}^2)} \qquad (8)$$

This means that if a laboratory uses the mass and skeletal density in determining the excess adsorption, then uncertainties will compound. If the skeletal density of a material is underestimated, then the assumption would bias the gravimetric excess adsorption measurements to be higher along with the calculated volumetric storage capacity. For example, the skeletal density of the commercially available activated carbon, MSC-30, is usually reported between 2.0

– 2.2 g/cm³. If its skeletal density is actually near 2.2 g/cm³ instead of the assumed 2.0 g/cm³, this would decrease the room temperature excess adsorption of hydrogen by about 5% at 100 bar and by about 7% at 200 bar.

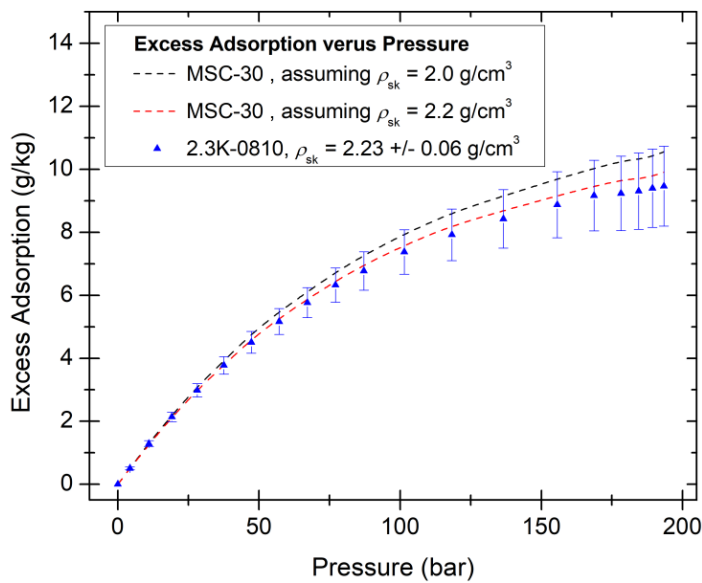

**Figure 4.** Hydrogen excess adsorption isotherm measurements of a coffee-based activated carbon and MSC-30. The measurements of MSC-30 were based on an assumed skeletal density of either 2.0 g/cm³ or 2.2 g/cm³. The error bars reported for 2.3K-0810 are based on the uncertainties in $V_d$, $V_r$, and $V_{sk}$.

Measurements of excess adsorption and storage capacities are highly sensitive to uncertainties in the skeletal volumes of the adsorbent material under test. Depending on the measurement system, these uncertainties compound when used in calculating derived quantities, such as volumetric storage capacities. It is recommended to measure the skeletal density of each material whenever possible. In cases where this is impractical, it is recommended to perform measurements in a well-calibrated, voluminous reactor cell on as large amount of the adsorbent material as possible to reduce the relative error.

## 5. Conclusions

We have presented measurements of the skeletal densities for a variety of adsorbent material types. The skeletal density is a key parameter in adsorbent material characterization. Measured skeletal densities of commercially available materials were consistent with the values reported in the cited literature. Skeletal densities of powdered activated carbon materials ranged from 2.09 – 2.54 g/cm³. Skeletal densities of non-exfoliated GCN materials ranged from 1.74 – 1.96 g/cm³ with an average density of 1.85 g/cm³. We have also demonstrated the critical role that accurate skeletal density measurements play in properly assessing a material's viability. If a skeletal density is underestimated by 10%, it can affect the room temperature excess adsorption by at least 5% at 100 bar and by 7% at 200 bar, depending on the material type.

## 6. ACKNOWLEDGEMENTS

The authors would like to thank the University of Missouri physics shop for their technical assistance and useful discussions. This work was supported by University of Missouri, and the U.S. Department of Energy under Contract No. DE-FG36-08GO18142. Z.A.P. acknowledges support from the National Science Foundation Graduate Research Fellowship under Grant Nos. 1540502 and 1451148. The identification of commercial products, contractors, and suppliers within this article are for informational purposes only, and do not imply endorsement by the University of Missouri, their associates, or their collaborators.

## 7. DATA AVAILABILITY

The data that support the findings of this study are available from the corresponding author upon reasonable request.

## 9. SUPPLEMENTARY MATERIALS

**NIST-traceable volume calibration spheres and silicon beads**

**Table 2:** Diameters and volumes of calibration spheres.

| Sphere Size | Part Number | Diameter [cm] | Volume [cm$^3$] | Uncertainty in Volume [cm$^3$] | Relative Uncertainty [%] |
|---|---|---|---|---|---|
| Medium | 01500-NIST-MED | 3.81 | 28.9583 | 0.0174 | 0.06 |
| Small | 01500-NIST-SM | 2.38 | 7.0699 | 0.0023 | 0.03 |
| Micro | 01500-NIST-MICRO | 1.27 | 1.0725 | 0.0006 | 0.06 |

NIST-traceable calibration volumes are available from vendors such as Quantachrome. They typically have a relative uncertainty of less than 0.07%. For applications in which obtaining NIST-traceable calibration spheres is either cost-prohibitive or not possible, non-adsorbing silicon beads may be used as a calibration displacement volume instead. Silicon beads (or "silicon shot") have a well-characterized density of 2.3290 g/cm$^3$ at standard temperature and pressure. Using the density with the dry mass yields the displacement volume.